Article

# Dynamic Versus Static Oxidation of Nb/Al-AlO$_x$/Nb Trilayer

**Tannaz Farrahi** [1*], **Alan W. Kleinsasser** [2], **Jie Wang**[1], **Michael Cyberey** [1], **Michael B. Eller**[1], **Jian Z. Zhang**[1], **Anthony R. Kerr** [3], **Joseph G. Lambert**[3], **Robert M. Weikle**[1], **Arthur W. Lichtenberger** [1]

1. Department of Electrical Engineering, University of Virginia, Charlottesville, VA, USA
2. Jet Propulsion Laboratory, California Institute of Technology, Pasadena, CA, USA
3. Central Development Laboratory, National Radio Astronomy Observatory, Charlottesville, VA, USA
* Correspondence: tf4xb@virginia.edu

**Abstract:** High-quality Nb-based superconductor-insulator-superconductor (SIS) junctions with aluminum-oxide (AlO$_x$) tunnel barriers grown from Al overlayers are broadly reported in the literature. However, thinner aluminum oxide tunnel barriers, i.e., with higher critical current density (J$_c$), yield significant leakage current in SIS junction I-V curves due to metallic point contacts in the tunnel barrier. It has been found that high quality, higher J$_c$ junctions can be realized with AlN barriers grown from Al overlayers; yet aluminum oxide-based barrier growth offers better control of J$_c$. Encouragingly, there have been recent reports of higher quality, high J$_c$ junctions from SIS AlO$_x$ tunnel barriers grown more slowly as well as dynamically (gas flowing) instead of statically. The goal of this work is to investigate the use of in situ ellipsometry in an attempt to develop a rapid materials analysis method to distinguish the different growth regimes responsible for the higher quality, high J$_c$ junctions. In this work, AlO$_x$ tunnel barriers are grown with 100% O$_2$ and 2% O$_2$ in Ar gas mixtures, both statically and dynamically, with in situ ellipsometry to measure the real-time barrier growth. Strikingly, different growth signatures are observed, dependent on the growth rate, the use of dynamic versus static oxidation and the dilution of the oxidation gas. This study presents arguments that associate these growth signatures with the point contact's material morphology and suggest further paths of investigation to realize even higher quality, higher J$_c$ AlO$_x$ barriers.

Nb/Al-AlO$_x$/Nb trilayer superconductor-insulator-superconductor (SIS) tunnel junctions are widely-used in the applications such as single flux quantum (SFQ), quantum bits (qubits), and millimeter and sub-millimeter wave heterodyne mixers [1–5]. Device performance is highly dependent on (J$_c$) and junction quality (e.g., the degree of subgap leakage), which in turn depends on the detailed nature of the ∼1 nm-thick insulating AlO$_x$ barrier material. The barrier thickness is expected to be controlled by the oxygen "Exposure" E = P$_{ox}$t$_{ox}$, where P$_{ox}$ is the oxygen partial pressure and t$_{ox}$ is the oxidation time. Historically, it has been found through junction electrical measurements that the AlO$_x$ has two growth regimes corresponding to "low" J$_c$ (< 10kA/cm2) and "high" J$_c$ tunnel barriers. High J$_c$ junctions have thinner AlO$_x$ layers, present increased leakage currents, and have a markedly stronger J$_c$ dependency on Exposure than low J$_c$ junctions [6–9]. This is generally explained by the existence of tunnel barrier defects (e.g., metallic /quantum point-contacts) formed in the initial oxide layer, which dominates high J$_c$ current transport through Multiple Andreev Reflections (MAR) in the superconducting state with high quantum-mechanical transparency [9–13]. The transition from high J$_c$ to the low J$_c$ growth modes is explained by the additional oxygen exposure's remedy of the metallic point-contacts and hence removal of these dominant parallel conduction paths for thicker barrier layers. The barrier growth in this thin barrier regime can therefore be thought as the in parallel growth of (a) the bulk of the oxide and (b) the forming and 'remedying' of the point-contacts.

High quality, high-$J_c$ junctions can be realized with AlN barriers grown by plasma nitridation of Al overlayers [14–20]; however, the AlO$_x$ barrier growth approach offers better control of $J_c$. Therefore, the superconducting community has a continued interest in developing and characterizing growth techniques of thinner oxide barriers with fewer defects. Historically, the growth of Gurvitch-style SIS trilayer [21] relies on static (no pumping of the oxidation gas) thermal oxidation of thin Al overlayers in pure oxygen for tunnel barrier formation. The fabrication processes have been extensively reported [6,7][22,23] including distinct low and high $J_c$ growth regimes [8]. There are fewer reports on dynamic oxidation (active flow of the oxidation gas due to pumping) and/or the use of diluted O$_2$. However, recent work by Tolpygo *et al.* [9][24] indicates that higher current densities can be realized using dynamic oxidation with either diluted O$_2$ gas mixtures or with 100% O$_2$ using orders of magnitude lower oxidation pressure, and hence significantly longer growth times, under proper chamber and Nb deposition conditions. This paper's work investigates four barrier growth modes in a single study for the first time: static (unpumped) and dynamic (pumped) oxidation using both 100% O$_2$ and diluted oxygen gas (2% O$_2$ in Ar, the largest dilution the gas manufacturer could repeatably provide) over a range of pressures, with all growths performed in the same oxidation chamber. In situ ellipsometry was used for the first time to measure and compare AlO$_x$ tunnel barrier growth in real time for these four oxidation modes. To the authors knowledge, this study used in situ ellipsometry for the first time to measure and compare AlO$_x$ tunnel barrier growth for Gurvitch AlO$_x$ barriers in general and for these four oxidation modes in particular. As discussed in section 2.1, all of the ellipsometer data in this paper was grown from thick Al (100 nm) on Si/SiO$_2$ wafers.

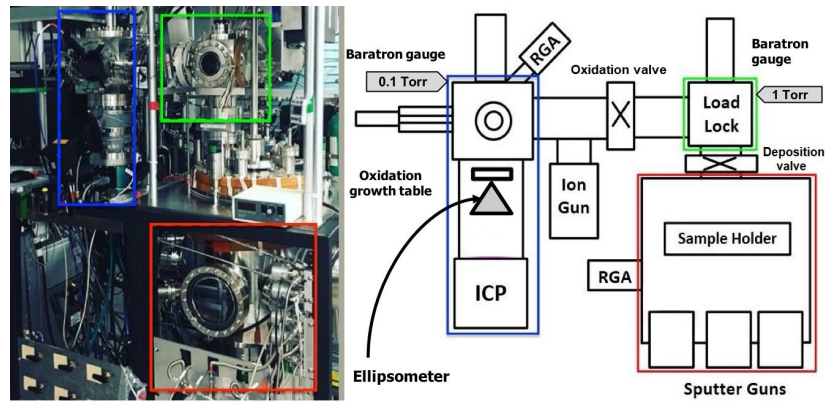

**Figure 1. Left**: Picture and **right**: schematic of the UVA trilayer system (Tri-3) modified for these experiments. Oxidation lines are connected to the load lock highlighted in green. The main deposition chamber and oxidation chamber are highlighted in red and blue, respectively.

1. Experimental Setup

The University of Virginia (UVA) trilayer deposition system (Tri-3), modified for these experiments, is composed of three chambers separated by two gate valves as shown in Figure 1. Three individual turbo-molecular pumps are used to evacuate the wafer entry load lock, main deposition chamber, and oxidation chamber (which is also used, in other experiments, for the nitridation of Al barriers). All three chambers are equipped with VAT Inc. pressure control valves. The load lock and oxidation chamber have 1 Torr (with a resolution of 0.1 mTorr) and 0.1 Torr (with a resolution of 0.05 mTorr) MKS Inc. Baratron gauges, respectively. The system base pressure is monitored using a residual gas analyzer (RGA). It has been determined for the UVA trilayer system that the tunnel barrier growth process is not influenced (as measurable by ellipsometer) by background oxidation of the Al overlayer once the base pressure of the system and the partial pressure of H$_2$O reach below 1E-8 and 3E-9 Torr (using different gauges), respectively [25]. The oxidation chamber is outfitted with an A. J. A. Woollam Co. M-2000U® ellipsometer with their Complete EASE software in order to record spectroscopic data, from 235-1000 nm at a fixed angle of 70°, during tunnel barrier growth. The ellipsometer uses 470 individual charge-coupled device (CCD) detectors to capture all wavelengths simultaneously for real time spectroscopic ellipsometry (SE) analysis across the entire spectrum.

Given that point-defects in the high-$J_c$ barrier material are of very small size as well as relatively low density (otherwise quasiparticle I-V characteristics would not be present) and the $AlO_x$ layer is only on order 1 nm thick, and additionally, high thickness-resolution material analytical instrument techniques will typically result in either the further oxidation or Ar ion bombardment of the revealed $AlO_x$ barrier, it is therefore challenging to chemically and/or spatially resolve and characterize such point contact defects. This work therefore does not attempt to model the ellipsometer's optical data differently for the low $J_c$ versus high $J_c$ material growth regimes. The optical data was fit using the Complete EASE software with the same approach authors previously reported for Al-AlN tunnel barrier growth [26]. Through the traditional use of fitted optical constants, the optically thin Al overlayer was modeled using a sum of Lorentz and a single Drude Oscillator, the Al-oxide layer was modeled using the customary Cauchy-Urbach dispersion relationships for weakly absorbing materials [27–29], and all optically thick Nb and Al films were modeled using pseudo optical constants obtained by directly inverting the ellipsometry equations in each measurement experiment from an ellipsometer scan of the given film. Small angular offsets from the substrate mounting can introduce appreciable run-to-run measurement errors. To correct for any angular offset in the system, the manufacturer provided an oxidized Si wafer with known thickness and material characteristics that was first measured before deposition and used as a calibration standard to correct for this offset. More details of the UVA Tri-3 system can be found elsewhere [30].

For this work, in order to investigate the growth of $AlO_x$ tunnel barriers, all of the oxidation experiments were performed in the separate oxidation chamber. The system base pressure and $H_2O$ partial pressure for all the reported experiments were ∼7.7E-9 and 1.4E-9 (or lower), respectively. All experiments used 50.8 mm diameter, 450 $\mu$m thick, double side polished Si substrates with 300 nm thermally grown $SiO_2$. During oxidation, the wafer, which is heatsunk with Apiezon® L grease to a metal wafer holder block, is lowered into a tapered opening of the 17 °C water-cooled growth table where the growth can be monitored in situ by the ellipsometer. The ellipsometer's light source module was aligned to just off the center of the wafer block during the initial system setup, and did not need to be further adjusted during this study. After positioning of the wafer-block on the oxidation growth table, and prior to introducing gas into the load lock and the start of oxidation, the ellipsometer is triggered to start acquiring data where each data point was acquired and integrated over ∼ 4 second intervals. In previous work, static oxidation of the $AlO_x$ tunnel barrier layer took place in the wafer load lock. For this work, the undiluted, static oxidation line of 99.995% purity $O_2$ was allowed to flow to the ellipsometer equipped oxidation chamber. In order to conduct this study's diluted oxidation experiments, a second oxidation line (dynamic line) was added to the wafer load lock. This line consists of a 2% $O_2$ in Ar gas source and a mass flow controller. For both static and dynamic oxidation, the gate valve between the load lock and oxidation chamber in Figure 1, called 'oxidation valve', is open and the gate valve between the load lock and main deposition chamber (called 'deposition valve') is closed. After growing Al (or Nb/Al) layers on a $Si/SiO_2$ substrate in the main chamber, the wafer is transferred to the growth table in the oxidation chamber without breaking the vacuum. In the case of static oxidation, prior to introducing oxygen gas to the system, both the load lock and oxidation chamber VAT valves are closed. For dynamic oxidation, the load lock VAT valve is closed and the oxidation gas pressure is controlled by adjusting the oxidation chamber VAT valve. For gas pressures below and above 90 mTorr, the 0.05 mTorr resolution oxidation chamber Baratron and the 0.1 mTorr resolution load lock pressure gauges are monitored, respectively. To achieve the desired working gas pressure in the dynamic oxidation setup for pressures below 90 mTorr, the oxidation chamber VAT valve position is adjusted automatically with feedback from the 0.05 mTorr resolution sensor, but for pressures above 90 mTorr, the VAT valve is adjusted manually. The oxidation gas flow is set at 28 sccm in both cases. After completing the oxidation and turning off the gas flow in both growth modes, in order to quickly evacuate the chambers of oxygen, both the load lock and oxidation VAT valves are immediately opened.

**2. Results and discussion**

*2.1. Initial Study-$AlO_x$ Thickness Repeatability*

In theory, the critical current density of a tunnel junction is exponentially dependent on the tunnel barrier thickness. Therefore, these experiments require a high degree of reproducibility in the ellipsometry measurement of barrier thickness. In an initial exploratory study, the repeatability of the ellipsometric-measured $AlO_x$ growth was assessed. Similar to the approach UVA uses for realizing trilayer for SIS junctions, after a "modest" ion gun clean to remove ∼10 nm of $SiO_2$ from the wafer surface, 165 nm of Nb followed by ∼5 nm Al overlayer were sputter deposited onto the wafer. Films were exposed to undiluted and diluted oxygen under both dynamic and static thermal oxidation settings for 8100 seconds and the thickness profile of each oxidation process was recorded in situ by the ellipsometer with a 25 µm by 60 µm spot size from just off the center of wafer. As described above, the optical constants of each Al overlayer film were modeled using a sum of four Lorentz and a single Drude oscillators [25,26].

A typical plot of thickness versus time has an initial period of rapid growth followed by a slower, decreasing growth rate. The average run to run thickness variation (beyond the initial rapid thickness rise) was found to be ∼0.04 nm. This variation in oxide thickness is significant when considered in terms of critical current density. The variation is believed to be primarily due to this study's ellipsometric growth model having to take into account both the changes in thin Al overlayer and underlying Nb thickness during the growth. Therefore, an optically opaque Al layer (∼100 nm thick) was instead use for the $AlO_x$ growth, and modeled using Cauchy and Urbach dispersion for the real and imaginary parts of the refractive index, respectively. The growth experiments were repeated with the same oxidation conditions on the 100 nm layer of aluminum deposited directly on $Si/SiO_2$ wafers, yielding a significantly more repeatable measured $AlO_x$ growth rate with only ±5 pm run to run scatter. Henceforth, unless otherwise specified, the ellipsometric data reported in this study are performed at room temperature and based on samples with a 100 nm aluminum layer. Since the small local variations in barrier thickness as well as point contact defects occur on an atomic scale, it is clear that the ∼1500 $\mu m^2$ sampling area effectively averages these local variations in barrier thickness. Since SIS junctions (which are typically on order $1 \mu m$ diameter) also averages local atomic scale variations in barrier thickness and morphology, it is reasonable to consider correlating ellipsometer measurements to SIS junction electrical characteristics (which in turn, is correlated in the literature to tunnel barrier morphology). Given the highly repeatable run to run ellipsometry outputted thickness, it was expected that the ellipsometer reported thickness would be a reasonable predictor of the relative $AlO_x$ barrier thickness. However, as will be detailed below, the data suggests that the ellipsometer reported thickness also reflects differences in the morphology of the barrier.

*2.2. Growth Versus Time*

This study first compared statically grown $AlO_x$ in 100% $O_2$ (the most commonly-used mode of trilayer growth) with dynamically grown $AlO_x$ in 2% $O_2$ at pressures from 3 mTorr to 100 mTorr. The results are shown in Figure 2a on a linear time scale. As expected, the oxide growth rate increases dramatically initially, and thicker layers are obtained with higher pressure in both cases. For a given pressure, as expected, the growth rate is significantly larger for the undiluted case. Based simply on Exposure considerations, in order to achieve a given oxide thickness for a given growth time, the $O_2$ partial pressure needs to be 50 times higher in the diluted case. The data is reasonably consistent with this expectation (for the only comparable growth runs) in the linear region with the 3 mTorr 100% O$_2$ data (3 mTorr partial pressure) occurring approximately 1.5x earlier in time than the 100 mTorr 2% O$_2$ (2 mTorr O$_2$ partial pressure) data.

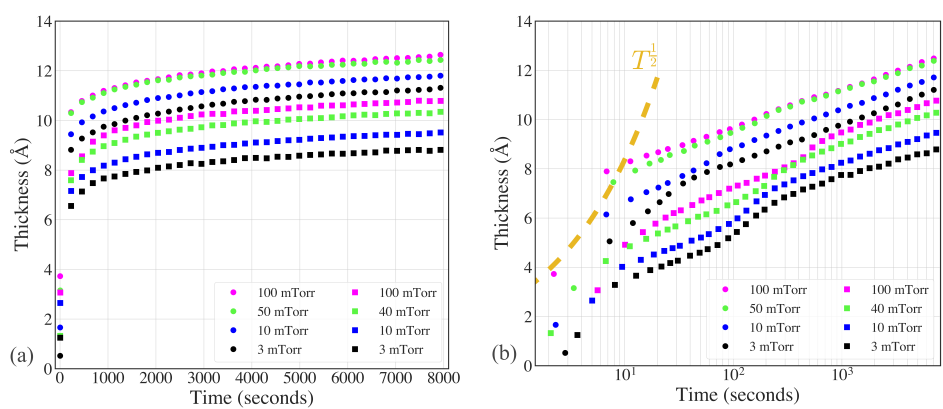

**Figure 2. (a)** The ellipsometer-outputted AlO$_x$ thickness as a function of time for dynamically grown oxide in 2% O$_2$ and statically grown oxide in 100% O$_2$ for different gas pressures. **(b)** The same data plotted as a function of time on log scale. Note that typically only two data points are contained in the first ten seconds of growth. The Mott-Cabrera theory ($L(t) \propto t^{1/2}$) is fitted for the first three 100 mTorr static experimental data points (first regime of AlO$_x$ growth) and is shown in the long-dashed orange curve.

The growth kinetics of oxides are described by the classic paper by Fromhold and Cook [31] based on the Mott-Cabrera [32] theory in which two regimes are defined for oxide thickness (L) as a function of time (t). This theory predicts that an initial thin layer of oxide film growth for which oxidation is extremely rapid. This behavior results from electrons tunnelling through the thin oxide (electron current is large), leading to the formation of an electronic potential on the film surface (the Mott potential). The oxide thickness in this first stage of growth is described by $L(t) \propto t^{1/2}$ (the Mott-Cabrera equation). In the second regime, oxide growth is limited by electron tunneling current through the thicker oxide without a potential on the surface of film. Growth in this stage is slow and logarithmic, described by $L(t)/\log(t)$ [32]. These two regimes are observed in this study's data as well as that of other groups [33–38] where in particular Lindmark [33],[38] also used in situ ellipsometric measurements of AlO$_x$ at different O$_2$ oxidation pressures, but for magnetic devices.

To further consider these results, the Figure 2a data is expanded and re-plotted on a logarithmic time scale as shown in Figure 2b. The sudden initial increase of thickness to several angstroms for all the curves presumably occurs in the first Mott growth regime where oxide forms on the metal surface through electron tunneling. As is evident from the plot, the first stage of growth occurs over the first 10 seconds or less of growth, while the growth curves become ∼linear on this log(t) plot for considerably longer growth times. Note that for all growths, the first logged ellipsometer data point does not correspond exactly to the moment that oxidation gas starts to flow. As described above, the ellipsometer is manually started before the introduction of the oxidation gas, and the ellipsometer takes data every 4-5 seconds. In plotting the data, authors have taken the last ellipsometer data point with no growth as t = 0. The result is a small inaccuracy due to the time shift of the Figure 2b curves. This error is of little consequence beyond ∼10 seconds. A fit of the Mott theory to the first three data points of the 100 mTorr, 100% O$_2$ static data (magenta, solid circles) is also shown, further illustrating the expected form of initial growth.

It is interesting to note that all of the oxidation curves in Figure 2b exhibit what this study describes as a third (middle) regime of growth, a modest, gradual inflection/step clearly above the initial ∼10 second first regime of growth, but below the eventual steady-state linear regime. For example, the 100 mTorr 2% $O_2$ dynamic data has such a step around 8-8.5 Å, while the 10 mTorr 100% $O_2$ dynamic data has a step around 6-7 Å. Similar step features were also found for our growth samples using Nb/Al (∼ 7 nm) layers. In fact, Lindmark *et al.* [33] reported a similar step in one of their data traces, but offered that it might have been due to a pressure fluctuation during growth. However, closer examination of their data actually reveals modest steps in more of their traces. Their static oxidation results were also marred by background gases as significant growth was seen even before the start of oxidation. They also used relatively short growth times such that meaningful plots of thickness versus Exposure cannot be constructed. It can be seen within the two sets of curves, (i) the inflection/step occurs earlier in time the lower the pressure (and hence, at lower exposure) for a given growth approach, and (ii) the step occurs over a longer time duration and with a larger height change, the lower the pressure. If the occurrence of these steps in time were strictly dependent on oxygen Exposure, the step for the 3 mTorr 100% $O_2$ static growth should occur 1.5x sooner than the 100 mTorr, 2% $O_2$ dynamic growth. However, the static step occurs at approximately a factor of ∼5 earlier. Additionally, the characteristics of these steps are quite different for the two growth sets where the inflection/steps are significantly more pronounced for 2% $O_2$ dynamic growth. This suggests that these two growth modes may have different underlying growth mechanics.

*2.3. all Four Growth Modes Versus Time*

Next, the ellipsometer outputted oxide thickness is plotted as a function of oxygen Exposure for all four growth modes. The results are shown in Figure 3a-d. Since $J_c$, theoretically, depends exponentially on $AlO_x$ thickness, we expect the Exposure dependence of the oxide thickness (on a linear scale) to resemble the $J_c$ dependence of Nb/Al-$AlO_x$/Nb tunnel junctions (on a log scale). The "third middle regime" step feature is exhibited in all four sub-figures of Figure 3, although the steps for undiluted dynamic growth are less pronounced (they are clearly visible when the data are blown up). For low to medium exposure values in all four growth modes, distinct thickness traces are found. In addition, the growth curves for the diluted dynamic, undiluted dynamic, and diluted static growth modes, but not the undiluted static mode (Figure 3d), coalesce after the step into a single consolidated dependence.

Consider Figure 3d, which compares the Exposure dependence of ellipsometer outputted thickness at various pressures for 100% static growth. The spread in ellipsometer outputted thickness for E = $10^4$ mTorr-sec across this pressure range is more than an angstrom. Yet this Exposure would realize SIS trilayer that is in the 'low' $J_c$ regime, without a significant number of point-contacts and hence would result in high quality electrical junction characteristics. Additionally, as has been described above, the $J_c$ of 100% static trilayer has been found to be uniquely determined by the Exposure value used in the oxidation growth for a wide range of oxidation pressures. Yet the > 1 Å range of ellipsometer outputted thickness would result in a $J_c$ variation significantly larger than what is experimentally found in the literature and from authors group's decades of experience fabricating SIS junctions. To understand this apparent contradiction, one must note that the thickness output of an ellipsometer is dependent not only on the actual oxide physical thickness, but also on the material characteristics of the oxide being measured and how well the ellipsometer model used is able to capture such nuanced characteristics. Note that all of the growth samples started with an aluminum layer deposited under identical conditions and hence differences in ellipsometer reported thickness (i.e., a change in measured optical constants) should be due to differences in the morphology of the $AlO_x$ from differences in the oxidation conditions. It is therefore this study's interpretation that the ellipsometer thickness data in this work is indicative of subtle differences in $AlO_x$ film morphology. In this light, the ellipsometer reported (non-physical) thickness spread of more than 1 Å in Figure 3d is now understood as indicating a pressure dependent (and hence likely a rate of growth dependence) subtle difference in $AlO_x$ morphology. The term "subtle" is used because the (argued) different $AlO_x$ morphologies would still yield the same $J_c$.

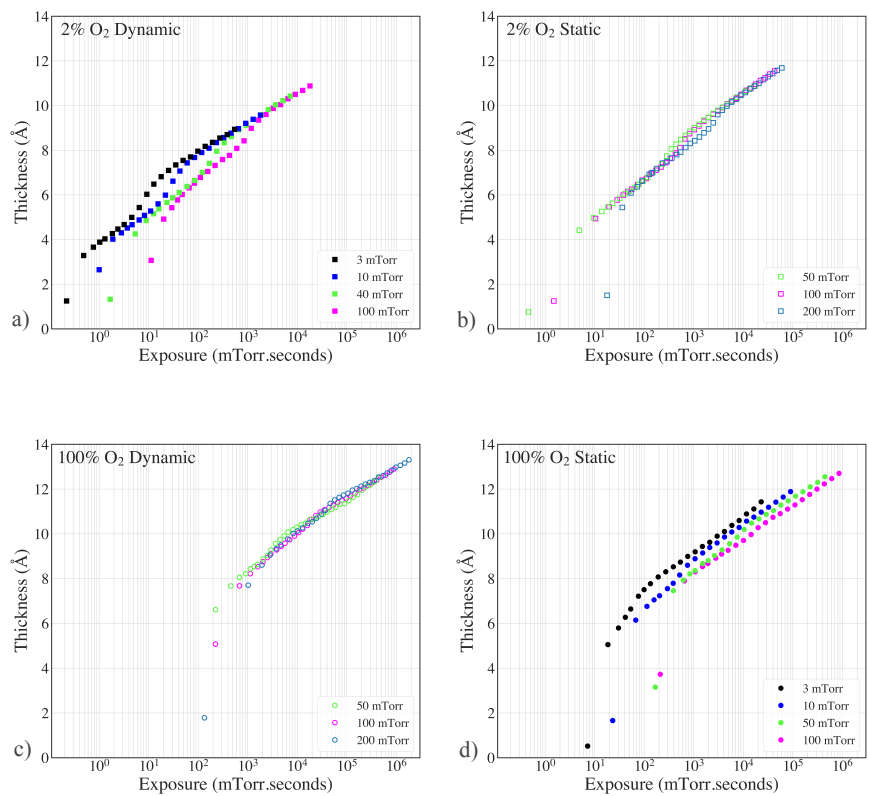

**Figure 3.** The ellipsometer outputted $AlO_x$ thickness as a function of Exposure for oxides grown (a) dynamically with 2% $O_2$, (b) statically with 2% $O_2$, (c) dynamically with 100% $O_2$, and (d) statically with a 100% $O_2$ for various oxygen pressures.

Given the spread of ellipsometer outputted thickness values in Figure 3d, one could attempt to adjust the ellipsometer model for each pressure to force the thickness outputs for high Exposure to one trace. However, based on the widely reported varied quality of measured SIS electrical characteristics on growth conditions and thickness, the morphology of the $AlO_x$ barrier must differ between the low and high $J_c$ regimes (i.e. with thickness), and also must vary depending how the barrier is grown (100% versus dilute $O_2$, and static versus dynamic). This study therefore purposefully choose to use the same ellipsometer model for all of the measurements of $AlO_x$ growth in this work in order to gain insight from the ellipsometer's material-characteristics influenced output. For example, as noted previously, in contrast to Figure 3d, the thickness curves in Figures 3a, b and c plots each coalesce to a single dependence. The structure of the thickness traces in Figures 3a, b and c are similar in characteristic, but more pronounced in Figure 3a. The differences in these sets of traces, and the differences in a given growth set of different pressures, are therefore arguably meaningful and can arguably tied to differences in material growth and the resulting electrical characteristics.

## 2.4. 2% $O_2$ Dynamic Growth

Consider Figure 3a, which compares the ellipsometer outputted thickness at various pressures for 2% $O_2$, dynamic growth. The inflection/step occurs (i.e., the coalescing to the consolidated trace) at a lower Exposure the lower the oxidation pressure. All of these steps occur at Exposures that are clearly in the high $J_c$ (thin barrier) regime. Given the lower the pressure the lower the growth rate dependency, authors therefore interpret the coalescing of a given trace (for this 2% $O_2$ dynamic growth mode) to the consolidated path as being connected to the 'fixing' of the point-contacts that are responsible for the recent reports of higher quality, high $J_c$ SIS junctions from slower oxidation growth. This data is consistent with the new understanding that a slower oxidation growth rate will 'fix' the point contacts at a lower Exposure. These results raise the question of how low of an $O_2$ partial pressure (i.e., how low of a growth rate) might still offer improved electrical characteristics and is there additionally perhaps a slow enough rate for which the density of point-contacts initially formed is significantly reduced. Since the coalescing of 3 mTorr trace occurs substantially below the 10 mTorr trace, further improvements in SIS electrical junction characteristics may arguably be found below the 3 mTorr 2% $O_2$ condition (0.06 mTorr $O_2$ partial pressure).

Note that Tolpygo in [9] used a dynamic growth condition of 0.6 mTorr $O_2$ partial pressure. The use of very low $O_2$ partial pressure conditions can present a practical challenge since the oxidation growth time to realize a desired current density could take days or longer. Wang [39] recently used a two-step oxidation approach, the first growth at low pressure (0.1 mTorr with undiluted $O_2$, giving an $O_2$ partial pressure of 10 mTorr) and the second growth at a significantly higher pressure, to realize a more manageable total growth time. Future trilayer system modifications are planned in order to explore such a two-step oxidation processes, with an envisioned first step control to as low as 0.01 mTorr (which, with 2% gas, would yield an $O_2$ partial pressure of 2E-4 mTorr, orders of magnitude lower than what is currently being explored by others). As part of this future study, the same in situ ellipsometer measurement approach will be used to determine if the coalescing ellipsometer signature saturates with Exposure below some critical oxidation pressure. If growth traces are found to saturate below a critical pressure, the study will explore whether this critical coalescing-pressure also marks the lowest oxidation pressure in a two-step (Wang) oxidation process to give further improvements for high $J_c$ electrical characteristics. This ellipsometer approach will also be used to help determine the duration to use for the first, low pressure oxidation step. Arguably, the first oxidation step should be grown to an Exposure well past the coalescing signature for the given growth pressure.

*2.5. 2% $O_2$ Dynamic Versus 100% $O_2$ Dynamic and 2% $O_2$ Static Growth*

Given that low partial pressure, dynamic growth has recently been used to realize improved electrical characteristics for higher critical current densities, this growth regime appears to be the most promising to further investigate and exploit. However, it is worth further consideration of our ellipsometer data for the other growth regimes. Since the four growth modes explored in this work are distinct, one should not expect the barrier growth for each mode to be identical and hence one should also not necessarily expect the ellipsometer measured trace characteristics for the four growth modes to be identical either.

The ellipsometer outputted $AlO_x$ film thickness series are plotted versus Exposure on one graph in Figure 4 for the three growth modes found in Figure 3a, b and c. For sufficiently high Exposure, the 100% $O_2$ dynamic and the 2% $O_2$ static traces coalesce to the single 2% $O_2$ dynamic trace. Moreover, as was discussed in section 2.4 for 2% $O_2$ dynamic growth, within each of the 100% $O_2$ dynamic and the 2% $O_2$ static growth modes, the coalescing to the consolidated trace occurs at a lower Exposure, the lower the pressure (i.e., the lower the growth rate). Since the general signature of the ellipsometer traces for 100% $O_2$ dynamic and 2% $O_2$ static is similar to the 2% $O_2$ dynamic ellipsometer signature, the data argues that there is a similarity in the growth mechanics of these three growth modes. However, the coalescing signature of the 100% $O_2$ dynamic and 2% $O_2$ static ellipsometer traces are much less pronounced. The author's original expectation of the role of diluted versus non diluted oxidation was simply that a diluted oxidation gas allowed one to obtain a slower oxidation rate for the same Exposure value. And the original expectation of the role of static versus dynamic oxidation is that dynamic oxidation reduces the potential effects on oxidation mechanics from chamber outgassing. With this logic, unless 100% O$_2$ and/or static conditions have additional unknown impacts on the growth mechanics, one would likely argue that the less pronounced coalescing signature for 100% dynamic growth is due to the faster growth from the larger O$_2$ partial pressure (the smallest 100% O$_2$ dynamic pressure used for Figure 4 gives an O$_2$ partial pressure of 50 mTorr, compared to the highest partial pressure used for 2% dynamic growth of 4 mTorr). And one would also likely argue that the less pronounced signature for 2% O$_2$ static growth is due to (since 2% O$_2$ is used for both cases) the background gases from chamber outgassing. However, as will be shown in section 2.6, there are arguably additional factors involved.

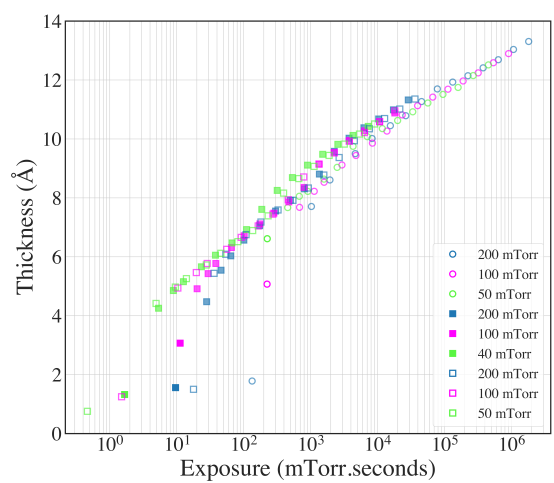

**Figure 4.** The measured AlO$_x$ thickness, as determined through spectroscopic ellipsometry, as a function of oxygen exposure is plotted for statically grown oxide in 2% O$_2$ (open squares), dynamically grown oxide in 2% O$_2$ (solid squares), and dynamically grown oxide in 100% O$_2$ (open circles).

*2.6. 2% O$_2$ Dynamic Dynamic Versus 100% Static Growth*

The ellipsometer outputted AlO$_x$ film thickness series are plotted versus Exposure on one graph in Figure 5a for the two growth modes found in Figure 3a and 3d (Figure 5b is an expanded view) for varying pressure. For low Exposure, distinct thickness traces for each pressure are present for both growth modes. In this low exposure regime, for a given Exposure, a lower pressure results in a larger outputted thickness for a given growth mode. Additionally, for low Exposures and for the same Exposure and pressure, the outputted dynamic oxide thickness is significantly larger than the static oxide thickness. Given author's interpretation of the thickness *output* of this study's ellipsometer being dependent not only on the *actual* oxide physical thickness, but also on the material characteristics of the oxide being measured, this dependency at low Exposure is arguably an indication of a higher quality barrier (fewer metallic/quantum point-contacts) the lower the pressure and the lower the O$_2$ partial pressure due to a slower oxidation rate. Note that a growth process that remedies the point-contacts at a thinner barrier layer thickness may also result in a smaller spread in barrier thickness.

For large Exposure, while the undiluted O$_2$ static growth traces do exhibit the modest step signature, in contrast to the diluted dynamic conditions (as well as the other two growth modes), these step signatures do not coalesce to a single consolidated dependence. Figure 5b is an expanded view of Figure 5a that clearly shows the coalescing of the diluted dynamic traces, in contrast to the case of the undiluted static growth where each pressure trace follows a different growth path, even for very large Exposure. For high Exposure, the 10 mTorr undiluted static growth trace coincidentally falls on the coalesced consolidated trace of the other growth modes. It is not clear why the undiluted static growth mode gives such different results from the three other growth modes. From section 2.5, this study's original expectation of the role of diluted versus non diluted oxidation was simply that of the impact on oxidation rate for given Exposure value. The original expectation of the role of static versus dynamic oxidation was simply that of impacts from chamber outgassing. With this logic, one would likely argue that the lack of coalescing to a consolidate trace for undiluted static growth is due to the faster growth and/or outgassing effects. However, both the diluted static and the undiluted dynamic growth modes exhibit coalescing to a consolidated trace at high Exposure. The distinctly different trace signatures of the undiluted static growth are therefore arguably due to some combined effect of high growth rate and background gases on AlO$_x$ growth, or additional individual impacts on the growth mechanics.

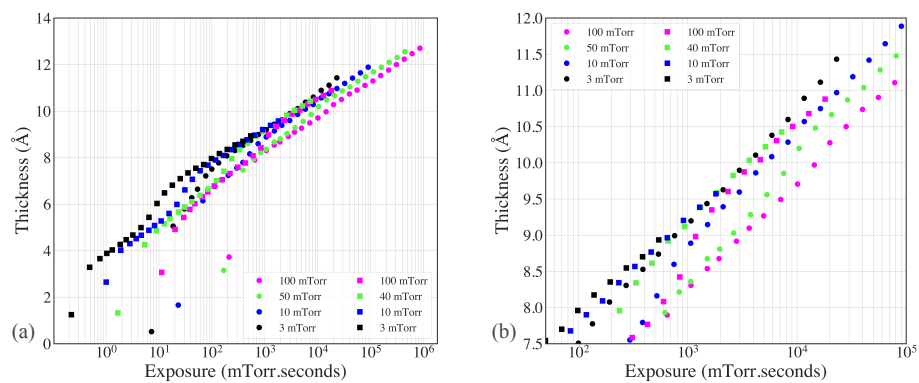

**Figure 5.** **Left: (a)** The outputted AlO$_x$ thickness, and **right: (b)** the expanded view of the measured AlO$_x$ thickness as determined through spectroscopic ellipsometry, as a function of oxygen Exposure (oxygen partial pressure times time) is plotted for dynamically grown oxides in 2% O$_2$ (solid squares) and statically grown oxides in 100% O$_2$ (solid circles) for various oxygen pressures.

*2.7. 2% O$_2$ Dynamic Versus 2% and 100% O$_2$ Static Growth*

The ellipsometer outputted AlO$_x$ film thickness is plotted versus time (log scale) in Figure 6 for select pressures of three growth modes found in Figure 3a, b and d. The lowest O$_2$ partial pressure (3 mTorr) trace for undiluted static growth is compared against other growth mode traces at higher pressures. For a given pressure, the ellipsometer outputted traces versus time are quite similar for the diluted static and diluted dynamic growth modes. This suggests that the growth rates and the growth mechanisms for diluted growth are not strongly dependent on the choice of static versus dynamic conditions (even though the coalescing signature is stronger for diluted dynamic growth). This is in contrast to the results shown in section 2.6, where arguably the growth rates and growth mechanisms for undiluted growth are dependent on the choice of static versus dynamic conditions. As described in section 2.8, test wafers with Al-oxide tunnel barriers grown in the low-J$_c$ regime using the diluted dynamic oxidation method were found to follow the same J$_c$ versus Exposure relationship as UVA's previous and longstanding undiluted static trilayer growth method. These results reinforce the above expectation that the growth rate for diluted growth is not dependent on the choice of static versus dynamic conditions. Additionally, the 3 mTorr undiluted (3 mTorr O$_2$ partial pressure) static trace at high Exposure is fairly similar to the 200 mTorr 2% diluted (4 mTorr O$_2$ partial pressure) dynamic and static consolidate traces. However, this is believed to be coincidental since the undiluted static growth traces do not coalesce to a consolidated trace at high Exposure.

*2.8. J$_c$(E) Versus Exposure*

A full investigation of the electrical characteristic of SIS trilayer grown in the four growth modes for different Exposures and pressures is beyond this current study. Addition-ally, significant changes to the deposition system are required in order to realize the desired 2E-4 mTorr control of the O$_2$ partial pressure of oxidation required for the envisioned future investigation. However, two trilayer test wafers with AlO$_x$ tunnel barriers were grown to verify whether the J$_c$ of this study's trilayer for diluted dynamic oxidation would roughly follow the existing J$_c$ versus Exposure relationship for undiluted static oxidation. The two 2% dynamic growth conditions were (a) 8180 seconds at 400 mTorr (i.e., E = 65400 mTor-sec), and (b) 9450 seconds at 40 mTorr (i.e., 7560 mTorr-sec). The UVA SIS junction fabrication process, that has been extensively described elsewhere [40,41], was used to fabricate SIS junctions. J$_c$ was determined from R$_n$A assuming I$_c$R$_n$ = 1.8 mV (the typical 4.2 K value of authors trilayer from prior work), where R$_n$ is the measured junction resistance (the ratio of junction V and I at 5 mV) and A is the measured junction area. The resulting J$_c$ values were found to be 3 and 10.7 $\frac{kA}{cm^2}$, respectively. These two experimental data points are shown in the Figure 7 on a log-log plot of J$_c$ versus E. The established UVA empirical AlO$_x$ formula for undiluted static growth with J$_c$ in the range of 1 to 7 $\frac{kA}{cm^2}$, is also shown with a black dotted line of the form of J$_c$ ∝ E$^{-\alpha}$, with $\alpha$ = 0.4. The fit for this work's data is shown as a green dashed line with $\alpha$ = 0.48 in this low J$_c$ regime. For reference, the dashed blue lines show the Kleinsasser et al.'s trend lines, based on averaged data of many groups, where $\alpha$, for low and high J$_c$ regions (predominantly undiluted, static oxidation), is 0.4 and 1.6, respectively [6–8]. The red solid lines display the trends from the recent work of Tolpygo et al.'s, where $\alpha$ for the high J$_c$ region with diluted dynamic growth is 1.0 and $\alpha$ for the undiluted static growth is 0.521 [9]. Authors note that this study's initial J$_c$ versus Exposure data for diluted dynamic growth is reasonable.

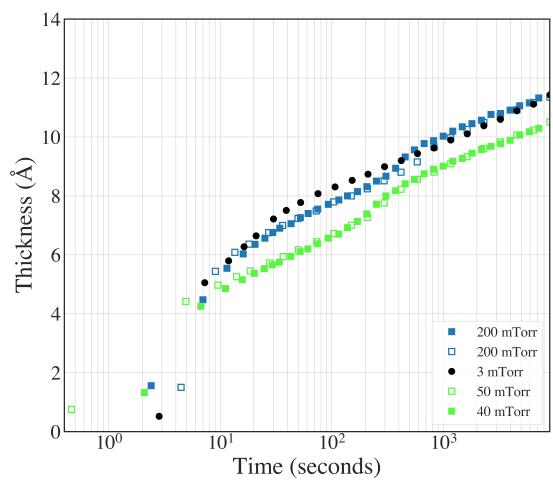

**Figure 6.** The measured AlO$_x$ thickness, as determined through spectroscopic ellipsometry, as a function of oxidation time is plotted for statically grown oxide in 2% O$_2$ (open squares), dynamically grown oxide in 2% O$_2$ (solid squares), and statically grown oxide in 100% O$_2$ (solid circles) for selected pressures.

## 3. Discussion

This work, studied AlO$_x$ growth with diluted (2% O$_2$) and undiluted (100% O$_2$) oxygen under both dynamic and static oxidation processes, using in situ spectroscopic ellipsometry. Several interesting results and arguments follow:

(i) The significant spread in ellipsometer outputted thickness for 100% static growth (the most used oxidation mode) for a given high Exposure condition across different pressures is much larger than physically reasonable since J$_c$ of such 100% static trilayer has been found to be instead uniquely determined by the Exposure value used in the oxidation growth for a wide range of oxidation pressures. All of the growth samples started with an aluminum layer deposited under identical conditions and hence differences in ellipsometer reported thickness (i.e., a change in measured optical constants) should be due to differences in the morphology of the $AlO_x$ from differences in the oxidation conditions. It is therefore authors interpretation that the ellipsometer thickness data in this work is also indicative of material differences in the morphology and potentially also the local thickness uniformity of the grown $AlO_x$ barrier layer.

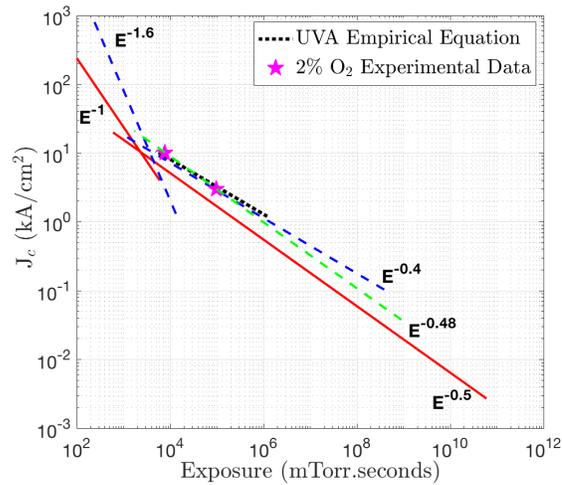

**Figure 7.** The measured $J_c$ data of SIS junctions with dynamically grown tunnel barrier in 2% $O_2$ as a function of oxygen Exposure. The $J_c$ experimental data are shown by solid magenta star markers. The dashed blue lines show the Kleinsasser *et al.* low and high $J_c$ regime while red solid lines display oxidation exponent trends for Tolpygo *et al.* $J_c$ regime [9]. The UVA empirical formula for the $AlO_x$ static growth in pure $O_2$ is shown in black dotted line for the $J_c$ range of 1 to 7 $\frac{kA}{cm^2}$. The best fit between the UVA experimental data is shown in green dashed line and has oxidation exponent of $E^{-0.48}$ in low $J_c$ regime.

(ii) In addition to the expected two 'standard' regimes of growth, all traces exhibit a third growth regime appearing as an intermediate gradual step. For a given growth approach, the step occurs at a lower Exposure the lower the pressure (i.e., the lower the growth rate). For a given growth approach, the step is also wider and taller the lower the pressure. This signature is significantly more pronounced for diluted (slower) dynamic growth than the three other growth modes. However, the ellipsometer outputted thickness traces versus time for a given pressure are quite similar for the diluted static and diluted dynamic growth modes. This suggests that the growth rates and the growth mechanisms for diluted growth are not strongly dependent on the choice of static versus dynamic conditions (even though the coalescing signature is stronger for diluted dynamic growth). Wang's recent work has provided experimental evidence in support of this understanding [39].

(iii) Based on the reported varied quality of measured SIS electrical characteristics on growth conditions and $AlO_x$ thickness, it is known that the morphology of grown $AlO_x$ barriers differ in the low versus the high $J_c$ regimes (i.e. with thickness), and also vary depending on how the barrier is grown (100% versus dilute $O_2$, and static versus dynamic). The same ellipsometer model was therefore purposefully used for all of the measurements of $AlO_x$ growth in this work in order to gain insight from the ellipsometer's material-characteristics influenced output.

(iv) For low Exposure, in all four growth modes, distinct thickness traces are found for each pressure, even for constant Exposure. Given this work's interpretation of the thickness output of ellipsometer being dependent not only on the actual oxide

    physical thickness, but also on the material characteristics and potentially also the local thickness uniformity of the oxide being measured, this dependency at low Exposure is arguably an indication of the quality of the AlO$_x$ barrier (connected to metallic/quantum point-contact growth).

(v)  The growth curves for three of the growth modes (diluted dynamic, undiluted dynamic, and diluted static) when plotted versus Exposure on a log scale coalesce to a single consolidated growth trace at sufficiently high Exposure. Individual growth traces for a given growth mode join the coalesce-line at lower Exposure the lower the growth pressure (i.e., the slower the growth rate and hence the longer the growth time). Slow dynamic growth has recently been found (Tolpygo[9], Wang[39]) to give improved SIS junction electrical characteristics for higher J$_c$. Considering the diluted (slow growth) dynamic growth mode further, and given the above considerations, this work interprets the coalescing of a given trace (for this 2% O$_2$ dynamic growth mode) to the consolidated path as being connected to the 'fixing' of the point-contacts for thinner barriers that is responsible for the recent reports of higher quality, high J$_c$ SIS junctions from slower oxidation growth. This work's data is therefore consistent with the new understanding that a slower oxidation growth rate will 'remedy' the point contacts at a lower Exposure. Given that all three of the above growth modes exhibit similar coalescing signatures and that both diluted modes arguably (point ii) have similar growth mechanisms, this interpretation of the coalescing to the consolidated trace being connected to the 'remedying' of the point-contacts may well apply to all three of these growth modes.

(vi)  However, for large Exposure, while the undiluted O$_2$ static growth traces do exhibit the modest step signature, in contrast to the other three growth modes, these step signatures do not coalesce to a single consolidated trace. One might argue that the lack of coalescing to a consolidate trace for undiluted static growth is due to the faster growth and/or outgassing effects. However, both the diluted static (which should entail the same level of outgassing) and the undiluted dynamic growth (which should entail the same large growth rates) modes exhibit coalescing to a consolidated trace at high Exposure. The different high Exposure signatures of the undiluted static growth may therefore be due a combined effect of both high growth rate and background outgassing gases on AlO$_x$ growth. While individual groups have reported realizing high electrical quality SIS junctions in the low J$_c$ regime with static oxidation performed at different pressures, authors are not aware of a study using the same trilayer system to compare the growth regimes of SIS junctions from AlO$_x$ barriers grown statically at significantly different oxidation pressures.

Future modifications of our trilayer system's oxidation chamber are planned to improve control of significantly lower oxidation pressures. Such changes are required in order to further leverage this study's in situ ellipsometer approach to investigate the following considerations: (A) Determine if the thickness coalescing signature saturates (thickness traces start to coincide) with Exposure below a critical oxidation pressure. If growth traces are found to saturate below a critical pressure, the future study will explore whether this critical coalescing-pressure also marks the lowest oxidation pressure in a two-step (Wang) oxidation process to provide further improvements for high J$_c$ electrical characteristics; And (B) Explore a two-step (Wang) oxidation processes, with an envisioned first step control to as low as 0.01 mTorr (which, with this study'd 2% gas, would yield an O$_2$ partial pressure of 2E-4 mTorr, orders of magnitude lower than what is currently being explored by others). The future study will also use the ellipsometric approach to determine the duration to use for the first, low pressure oxidation step. Arguably, the first oxidation step should be grown to an Exposure well past the coalescing signature for the given growth pressure.